# How Do Space-Time Digital Metasurfaces Serve to Perform Analog Signal Processing?

*Hamid Rajabalipanah[1], Ali Abdolali[\*,1], Shahid Iqbal[2], Lei Zhang[2], Tie Jun Cui[2]*

[1]Applied Electromagnetic Laboratory, School of Electrical Engineering, Iran University of science & Technology, Tehran, Iran (abdolali@iust.ac.ir).

[2] School of Information Science and Engineering, State Key Laboratory of Millimeter Waves, Southeast University, Nanjing 210096, China

ABSTRACT. In the quest to realize analog signal processing using sub-wavelength metasurfaces, in this paper, we demonstrate the first experimental demonstration of programmable time-modulated metasurface processors based on the key properties of spatial Fourier transformation. Exploiting space-time coding strategy enables local, independent, and real-time engineering of not only amplitude but also phase profile of the contributing reflective digital meta-atoms at both central and harmonic frequencies. Several illustrative examples are demonstrated to show that the proposed multifunctional calculus metasurface is capable of implementing a large class of useful mathematical operators, including $1^{st}$– and $2^{nd}$–order spatial differentiation, $1^{st}$–order spatial integration, and integro-differential equation solving accompanied by frequency conversions. Unlike the recent proposals, the designed time-modulated signal processor effectively operates for input signals containing wide spatial frequency bandwidths with an acceptable gain level. Proof-of-principle simulations are also reported along with the successful realization of image processing functions like edge detection. This time-varying wave-based computing system can set the direction for future developments of programmable metasurfaces with highly promising applications in ultrafast equation solving, real-time and continuous signal processing, and imaging.

## 1. Introduction

The history of analog computation comes from a number of electronic and mechanical computing machines developed to implement simple mathematical operations [1]. Due to having large sizes and slow responses, such analog computing devices were then totally overshadowed by the emergence of faster and more efficient digital integrated circuits in the second half of the 20[th] century, where a far superior performance for real-time analysis or processing applications was demanded [2]. Relying on a decade of fruitful development and the recent breakthrough in the seemingly unrelated field of metamaterials, *Silva et al.* [3] brought the analog computations back to the competition as "computational metamaterials" to overcome the speed and energy limitations as well as data conversion loss of digital techniques. In this way, different mathematical operations (spatial differentiation, integration, or convolution) can be realized as electromagnetic (EM) waves propagate through the metamaterial layer. Motivated by the recent renewed interest in wave-based analog signal processing, different proposals within two separate categories have been investigated in one of which the mathematical operators are directly realized in the real-space coordinate using a specially designed structure (Green's function approach in the literature) [4]–[14] and in the other group, the transfer functions associated with the operator of choice are realized in the spatial Fourier domain governed by graded-index lenses (metasurface approach in the literature) [15]–[18]. Although the former case offers great simplicity in accomplishing analog signal processing, it predominantly suffers from two major drawbacks of narrow spatial bandwidths and poor gain levels [6], [19]. This issue possesses great restrictions on the applicability of the architectures achieved by the second method in real-life scenarios. Besides, in a plethora of applications demanding real-time processing of signals and/or images, such as medical and satellite applications, reprogrammable and switchable accessing to different mathematical operators at the same time would bring tremendous benefits [20], [21]. Among them, spatial differentiation is a fundamental mathematical operation used in any field of science or engineering. In image processing, spatial differentiation enables image sharpening and edge-based segmentation, with broad applications ranging from microscopy and medical imaging to industrial inspection and object detection [22]. Although numerous efforts have been paid to expand the functionalities of wave-based signal processing systems, to the best of authors' knowledge, no programmable analog computing system has been reported. Most importantly, using the previous proposals based on the metasurface approach, an unwanted polarization conversion was forced to the analog computing system [16]. The present authors believe that the rapid developments in the land of programmable metasurfaces in recent couple of years provide sufficient maturity for them to resolve the unsolved challenges in this research domain.

Metasurfaces are defined broadly as artificial thin films created from subwavelength arrays of structured elements [23], [24] engineered to manipulate different properties of EM waves like polarization [25], amplitude [26], and phase [27]–[31]. Recently, a new class of metasurfaces called digital coding and programmable metasurfaces was pioneered by *Cui et al.* [32], which are represented in a digital manner with binary codes. The digital description of coding metasurfaces not only drastically facilitates and accelerates the related design and optimization procedures [33]–[41] but also is inherently compatible with switchable active elements, such as PIN-diodes [42], [43]. Hence, all coding elements of a digital coding metasurface can be independently controlled by a field-programmable gate array (FPGA). Through altering the coding sequences stored in the FPGA and then modulating coding meta-atoms, diverse EM functionalities can be switched in real time, thereby leading to programmable metasurfaces [44], [45]. However, in most of the recent studies, the coding sequences are generally fixed in time [46]–[49], and are changed by the control system only to switch the functionalities whenever needed (called as space coding). Although several types of research have focused on designing time-modulated metasurfaces, these approaches were based on analog modulations [50]–[52]. More recently, *Zhao et al.* introduced the first time-domain digital metasurface that enables efficient manipulation of spectral harmonic distribution [53]. Afterwards, different studies have been conducted to extend the arsenal of metasurface-based wave manipulations where a set of coding sequences are switched cyclically in a predesigned time period, enabling simultaneous manipulations of EM waves in both space and frequency domains [54]–[58]. It is worth noting that most of the above space-time-modulated metasurfaces serve to modulate only the phase profile of EM waves, leaving the space free for the other functionalities requiring both phase and amplitude modulations such as wave-based signal processing. To boost up the functionalities accompanied with the current analog computing systems, we will analyze the space-time coding metasurfaces from the perspective of reprogrammable analog calculus systems for real-time and parallel continuous data processing to answer this question that "*Are space-time digital metasurfaces useful to perform analog signal processing?*" .

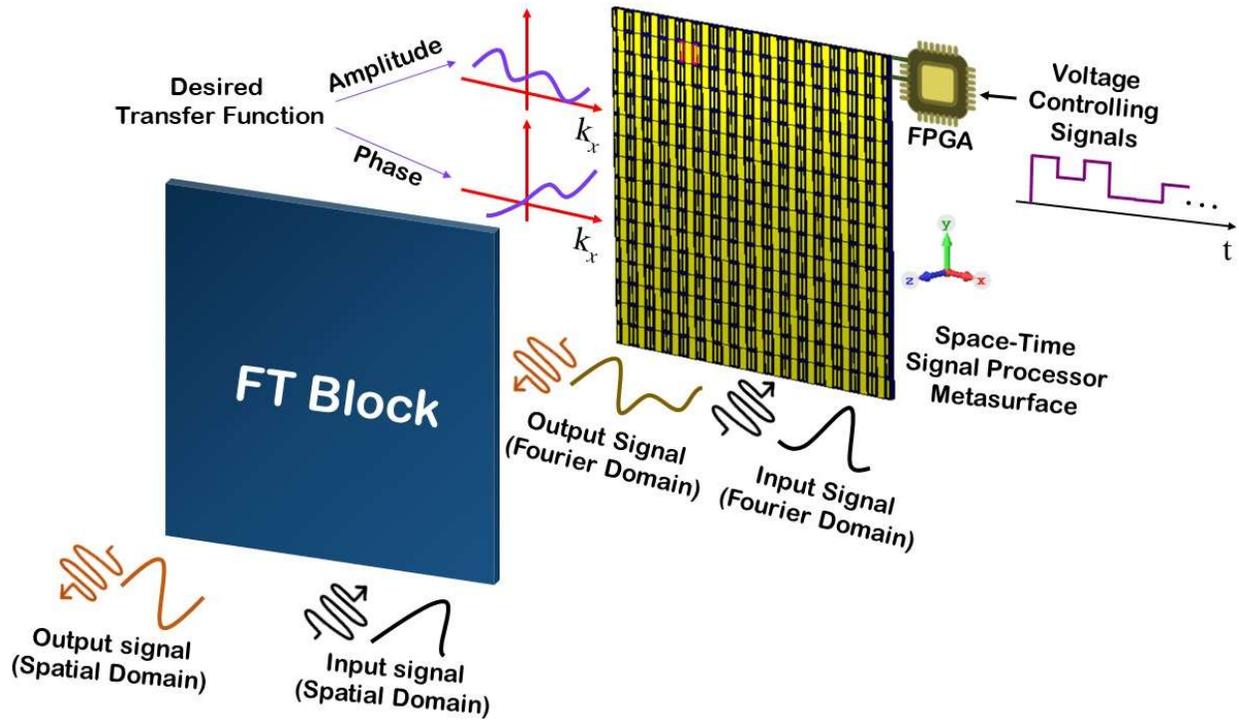

**Figure 1.** Schematic demonstration of the proposed analog signal processing scheme. The space-time digital metasurface serves to realize the transfer function associated with the operator of choice in the Fourier domain. The low-frequency digital signals are utilized to modulate the metasurface in order to instantaneously control the equivalent phase and amplitude of the occupying elements at a specific frequency harmonic.

In this paper, we propose a space-time digital metasurface as a reprogrammable versatile signal processor to realize different transfer functions such as 1st- and 2nd-order spatial differentiation, 1st-order spatial integration, and integro-differential equation solving based on the key properties of spatial Fourier transformation (metasurface approach). The space-time coding strategy is exploited to tailor the position-dependent phase and amplitude responses of the meta-atoms that imitate the transfer function associated with the operator of choice at both central and harmonic frequencies. To the best of our knowledge, it is the first time to realize multiple analog signal processing functions by using a single programmable architecture at microwave frequencies. In addition, we exploited the designed space-time coding metasurface along with a Fourier transformation layer to detect the edges for sharp changes in the incident field.

## 2. Analog Signal Processing Framework

The general concept of wave-based signal processing is graphically elucidated in **Figure 1**. The proposed sketch consists of two cascaded sub-modules: 1) a single a graded-index lens (or focusing metasurface) [59] to perform both Fourier and inverse Fourier transformation and 2) a reflective space-time metasurface to execute signal processing operators. According to the fact that FT{FT[$f$ (x)]}∝$f$ (−x), a single bi-functional block is enough to accomplish both FT and inverse FT transformations for the signals propagating along –z and +z directions, respectively, instead of using an additional block for IFT realization in the output [16]. Assuming a linear space invariant system in which $f(x)$ and $g(x)$ indicate the transverse field profiles pertaining to an arbitrary input signal and the corresponding output, respectively, the overall response can be theoretically modeled as:

$$g\left(x,y\right)=f\left(x,y\right)*h\left(x,y\right)=\iint h\left(x-x',y-y'\right)f\left(x',y'\right)dx'dy'.\qquad(1)$$

Here, $h(x)$ is the desired two-dimensional (2D) impulse response of the system and * stands for the linear convolution operation. Relying on the Fourier-transforming property of the Green lenses, the transverse distribution of the input wave will be translated into the spatial Fourier domain once it leaves the lens. Therefore, the mathematical formulation of the problem must be followed in the Fourier domain as:

$$G\left(k_x,k_y\right)=F\left(k_x,k_y\right)*H\left(k_x,k_y\right)\qquad(2)$$

in which, $G(k_x,k_y)$, $F(k_x,k_y)$, and $H(k_x,k_y)$ refer to the Fourier version of the output signal, input signal, and the transfer function associated with the desired signal processing operation, respectively, with $(k_x,k_y)$ being the 2D spatial frequency variables. Although realizing 2D mathematical operations can be obtained through modulating the metasurface along both x and y directions, without loss of generality, we limit our discussion to the 1D system which is symmetric along the y axis. The output field, in this case, is calculated by:

$$g\left(x\right)=FT^{-1}\left[\Gamma\left(x\right)F\left(k_x\right)\right]\qquad(3)$$

wherein, FT$^{-1}$ denotes the inverse Fourier transform. Note that the real-space coordinates $x$ at the metasurface represent $k_x$ as if the position-dependent reflection coefficient $\Gamma(x)$ implements the desired transfer function $H(k_x)$. Indeed, the real-space coordinate $x$ plays the role of $k_x$. Furthermore, any arbitrary transfer function with specific values over different spatial frequencies, $k_x$, can be elaborately realized by modulation of both reflection phase and amplitude responses of the proposed metasurface at the corresponding position, $x$.

## 3. Space-Time Digital Metasurface Design
### 3.1. Meta-atom Design

In this section, we exploited a 2-bit programmable digital particle, as illustrated in **Figure 2a,** to build the space-time coding metasurface realizing the above-mentioned signal processing concept. The coding element consists of three parts: 1) a metal along with two biasing lines, 2) a F4B substrate with $\varepsilon_r$=2.65, $tan\delta$=0.001, $h$=3 mm, and 3) a copper ground plane with the conductivity of $\sigma$=5.7×10$^7$ S/m to make a reflective structure [60]. The spatial periodicity of elements in both horizontal and vertical directions is chosen as $D$=12 mm so that the size of supercells comprising 7×7 meta-atoms becomes equal to a half wavelength at 3.5 GHz. As displayed in **Figure 2a,** two varactors are loaded in the digital meta-atom to connect the metallic strips together through a series RLC circuit. Two AC voltages are utilized to tune the capacitance level of the varactors, independently. We employed a series RLC model (R = 0.8 Ω, L = 0.7 nH, and controllable C) to circuitally represent the varactors in our full-wave simulations around the central frequency. A comprehensive parametric study has been accomplished to seek for the best group of four meta-atoms exhibiting constant $\pi/2$ phase differences when different voltage levels have been applied to the varactors. Thus, different capacitances of $(C_1, C_2)$=(2.7 pF, 2.7 pF), $(C_1, C_2)$= (2.7 pF, 0.7 pF), $(C_1, C_2)$ = (1 pF, 1 pF), and $(C_1, C_2)$=(0.6 pF, 0.6 pF) have been attained to elucidate the reflection response of "00", "01", "10", and "11" coding states, respectively. The other structural parameters are $L$=10 mm, $s$=5 mm, $w$=2 mm, and $g$=1.2 mm. To inspect the performance of the designed coding meta-atom, numerical simulations are executed via the commercial software package, CST Microwave Studio. Periodic boundary conditions are applied in the x and y directions while Floquet ports are assigned to the z-direction. An x-polarized plane wave normally shines on the meta-atoms. For different values of the varactor capacitance, the simulated reflection phases are illustrated in **Figure 2b**. As can be seen, a constant 90° phase difference with 0.2 GHz

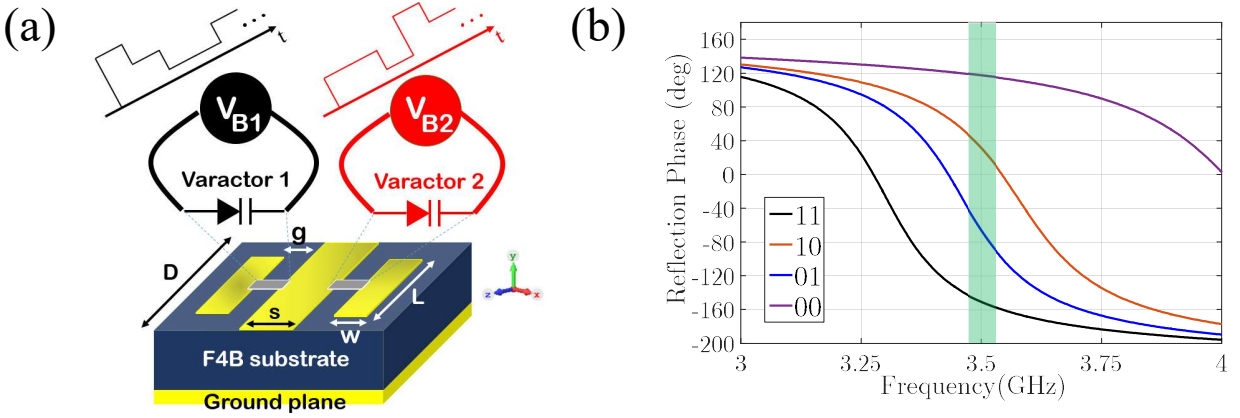

frequency bandwidth (3.4 GHz to 3.6 GHz) has been successfully achieved.

### 3.2. Time varying coding strategy

To realize the metasurface processor depicted in **Figure 1**, a space-time digital metasurface including sixteen programmable columns is utilized in which the columns are occupied with eight coding elements (**Figure 3**). The meta-atoms of each column are electrically connected by two biasing lines and hence, imitate identical digital code through sharing common time-varying control voltages. The overall size of the 1D metasurface processor is $680 \times 340$ mm ($8\lambda_0 \times 4\lambda_0$). In the framework of the space-time coding strategy, as shown in **Figure 3**, through applying a proper set of time-varying biasing signals, the digital code of each column cycles arbitrarily between 2-bit modes with a certain time periodicity, $T_m$ ($f_m=1/T_m$). Actually, the digital code of the meta-atoms can be potentially altered $L$ times in each time period in which the order of changes is expressed by a time-coding sequence, e.g., $\{10, 11, 11, 00\}$ with L=4. We should remark that the modulation frequency, $f_m$, is assumed much smaller than the incident wave frequency [61], [62]. According to the time-switched array theory and for a normal monochromatic plane wave, the meta-atoms are subject to a periodic time modulation of the reflection coefficient which can be mathematically expressed as $E_{ref,i}(t) = \Gamma_i(t) * E_{inc,i}(t)$ in which $\Gamma_i(t) = \sum \Gamma_i^n \Pi^n(t)$. Here, $\Gamma_i^n \epsilon \{e^{j0}, e^{j\pi/2}, e^{j\pi}, e^{j3\pi/2}\}$ refers to the reflection coefficient of the digital meta-atoms in the $i^{th}$ column and $n^{th}$ time interval (1≤n≤L), and finally, $\Pi^n(t)$ is a shifted pulse function with the modulation frequency of $f_m$ which possesses a non-zero value only during the $n^{th}$ interval. Fourier representation of the time-modulation reflection coefficients yields $\Gamma_i(t) = \sum a_i^m e^{j2\pi m f_m t}$ and $\Gamma_i(f) = \sum a_i^m \delta[2\pi(f - mf_m)]$ in which [54], [55]

$$a_i^m = \sum_{n=1}^{L} \frac{\Gamma_i^n}{\pi m} \sin\left(\frac{\pi m}{L}\right) \exp\left[-j(2n-1)\frac{\pi m}{L}\right] \qquad (4)$$

has been obtained after several mathematical manipulations. Theoretically speaking, $a_i^m$ discloses the equivalent reflection coefficient of the $i^{th}$ column at the $m^{th}$ frequency harmonic, i.e., $f+mf_m$.

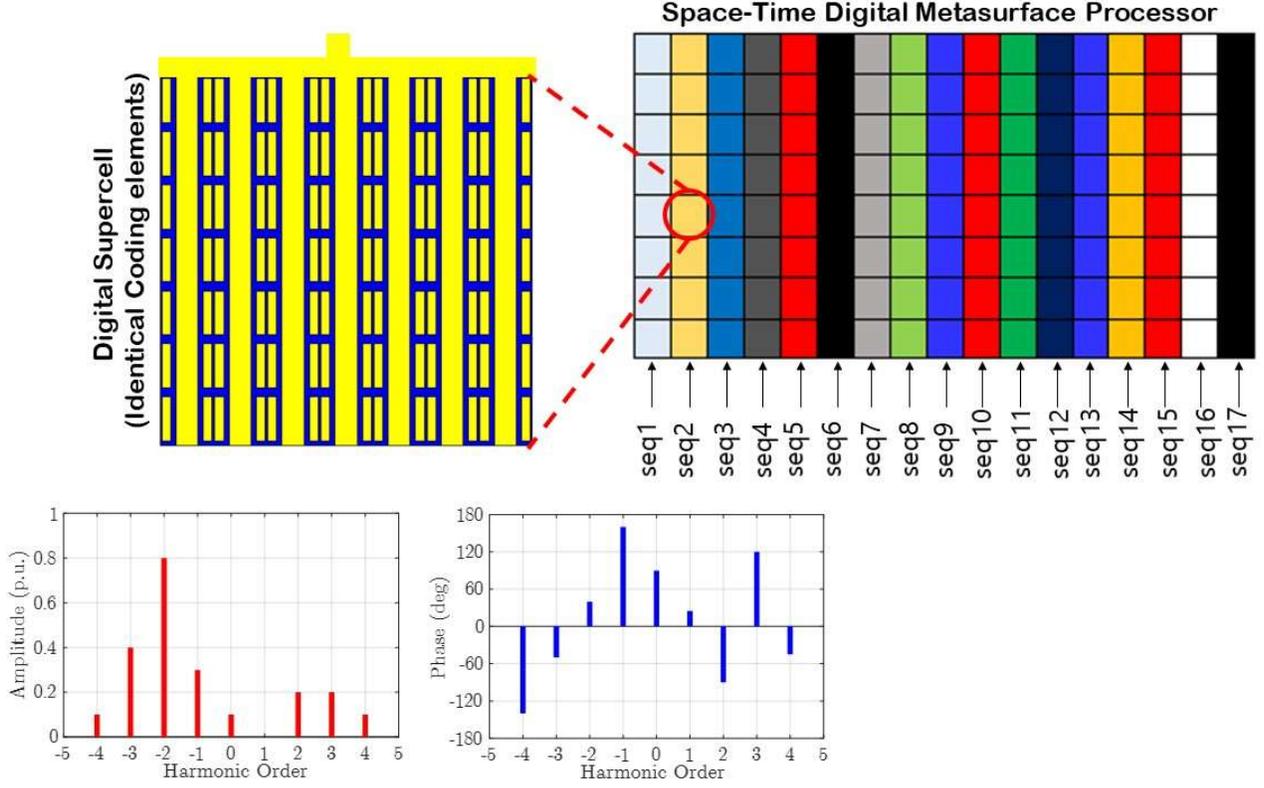

**Figure 3.** The overall scheme of the proposed space-time coding metasurface and its occupying supercells including 7×7 connected elements with identical digital states to implement the desired amplitudes/phases distribution dictated by the transfer function of interest, where different time-varying sequences lead to a specific harmonic amplitudes/phases distribution.

Being as a weighted average of the physical reflection coefficients, $a_i^m$ values drastically expand the range of our choices where arbitrary amplitudes and phases with high quantization levels can be artificially synthesized by using proper sets of time-coding sequences in spite of exploiting unitary-amplitude physical mete-atoms with only 2-bit phase modulation. This can be though as the inflection point of our study which elaborately enables analog signal processing with the digital programmable metasurfaces, an apparent paradox that can be resolved using space-time coding strategy. A larger time-coding sequence (larger L) results in more steps of available equivalent phase and amplitude in each harmonic. To prove the claim, let us consider the time-coding sequence of {00, 10, 10, 01, 11, 11, 11, 11} with L=8. Based on Eq. (4), the equivalent reflection coefficients for m=0 and m=−2 harmonics can be simply obtained as $0.4e^{-j10°}$ and $0.35e^{+j26.6°}$, respectively.

## 4. Results and Discussion

For our proposed system illustrated in **Figure 1**, two 17×8 and 25×8 reprogrammable metasurface processors with the dimension of $P_x \times P_y$ ($-P_x/2 < x < P_x/2$ and $-P_y/2 < y < P_y/2$) are responsible for realizing the positon-dependent amplitudes and phases corresponding to the desired operators of choice. In this way, the reflection coefficient of the time-modulated digital meta-atoms should spatially mimic the spectral trend of the transfer function associated with the desired mathematical operation. To show the versatility of the proposed signal processing system, the 1st-order and 2nd-order spatial differentiation, 1st-order spatial integration, and integro-differential equation solving are demonstrated in this paper. According to the Fourier transform principles, the transfer function accompanied by the above-mentioned 1D operators are

$$H\left(k_x\right) = j2k_x/P_x \tag{5}$$

$$H\left(k_x\right) = -4k_x^2/P_x^2 \tag{6}$$

$$H\left(k_x\right) = \begin{cases} 1 & |k_x| \leq d \\ d/jk_x & |k_x| > d \end{cases} \tag{7}$$

$$H\left(k_x\right) = \frac{jk_x}{\beta + j\alpha k_x - k_x^2} \tag{8}$$

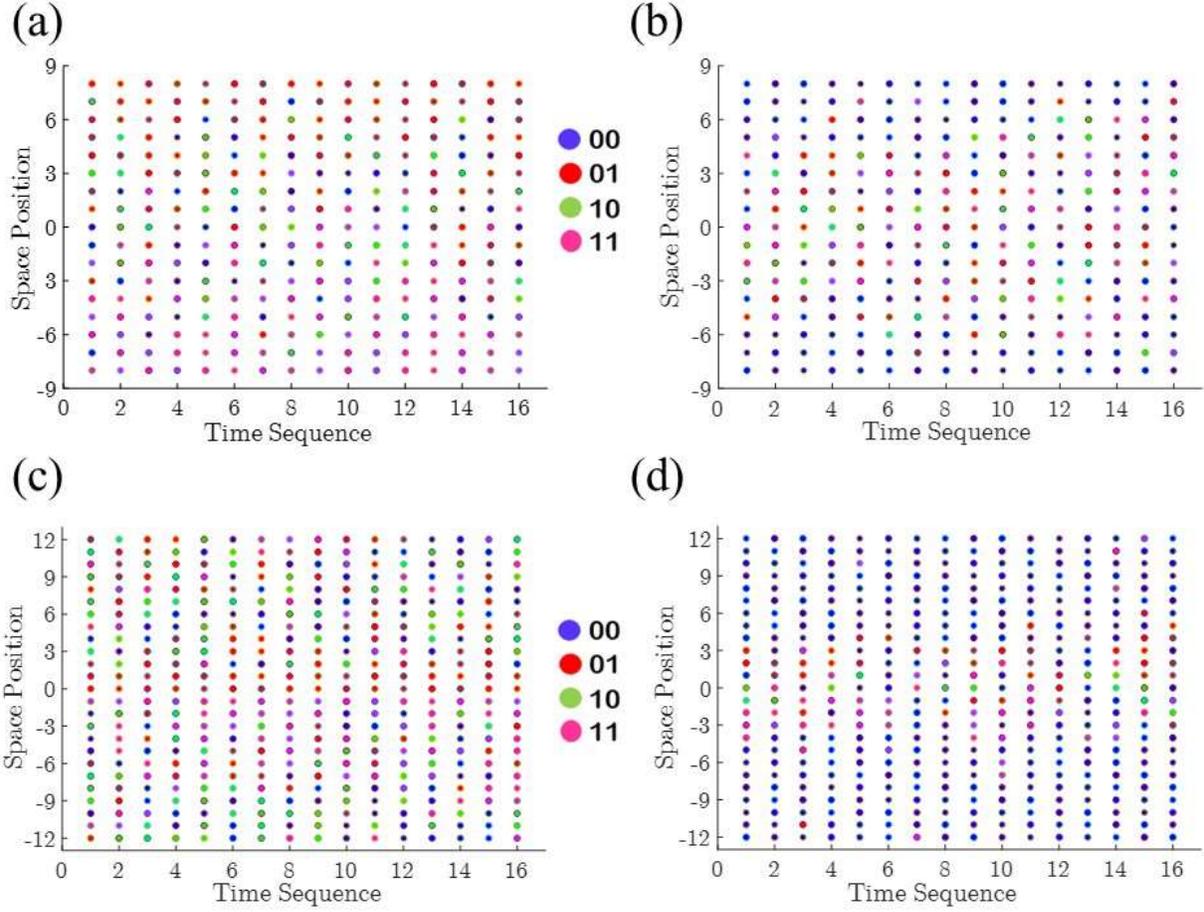

**Figure 4.** The optimum 1D space-time coding matrices for performing (a) 1st-order spatial differentiation, (b) 2nd-order spatial differentiation, (c) 1st-order spatial integration, and solving (d) integro-differential equation.

, respectively. The last transfer function belongs to the solving operator of a constant-coefficient ordinary integro-differential equation (CIDE), i.e., $dg(x)/dx + \alpha g(x) + \beta \int g(x)dx = f(x)$. Since the metasurface processor is inherently passive with a finite lateral dimension, we have normalized all transfer functions to guarantee that the maximum reflection coefficient of digital meta-atoms across the space-time metasurface is below unity and avoid gain requirements. Meanwhile, in the case of 1D integration, the singularity of the ideal transfer function at $k_x=0$ has been circumvented by truncating it within a spatial bandwidth of $2d=P_x/8$ in the vicinity of the zero harmonic.

In an attempt to realize Eqs. (5)–(8) with space-time digital metasurfaces at a specific frequency harmonic, the 2D space-time coding matrix of each sample is optimized so as to produce the equivalent phase and amplitude of the position-dependent reflection coefficient in the real-space coordinate, x, playing the role of the transfer function of choice in the spatial Fourier space, $k_x$. Although Eqs. (5)–(8) give seemingly continuous reflection profiles, the proposed calculus metasurfaces can tailor the scattered fields in integer steps of P, corresponding to the positions of the digital meta-atoms. In this way, four different space-time coding metasurfaces are programmed to implement the operators pertaining to the 1st-order and 2nd-order spatial differentiation, 1st-order spatial integration, and integro-differential equation solving. To present sufficient degrees of freedom, each column of the metasurface processors is modulated by suitable time-varying sequences with the length $L=16$. The frequency of modulation is chosen as 500 KHz [54] throughout the paper, which is much smaller than the center frequency. The corresponding space-time coding matrices set for the zeroth frequency harmonic are shown in **Figures 4a-d**. A full-wave analysis is carried out in CST Microwave Studio to authenticate the transfer function accompanied with each digital metasurface processor wherein an x-polarized plane wave impinges normally on the structure and the reflected electric fields along the transverse plane are plotted in **Figures 5-8**. Without loss of generality, the parameters of $\alpha$ and $\beta$ are chosen as 0.9 and 0.012, respectively. The amplitude and phase of the transfer functions are plotted in the Panels (a) and (c) of these figures, respectively. An excellent agreement between the numerical simulations and the ideal transfer functions depicted in the same figures proves the validity and versatility of the proposed signal processing approach. The linear and quadratic behaviors of the 1st-order and 2nd-order differentiation operations, respectively, are quite observable. It should be noted that the reset power is coupled to the other frequency harmonics. We emphasize that the space-time coding metasurface can be programmed to realize the transfer functions associated with different signal processing operators at the same time without adopting any change in the shape and geometry of the structure. Moreover, we should remark that the acceptable agreement between the results can be observed across the entire spatial frequency bandwidth, i.e., $|k_x| \leq k_0$, meaning that unlike analog computing platforms based on Greens' function approach [4]–[7], the space-time signal processors in this paper operate effectively even for those input signals containing a wide range of spatial frequency harmonics. In order to assess the quality of the output signals, the electric fields $f(x) = \exp(-x^2/c_1^2)$, $f(x) = sinc(x/c_2)$, $f(x) = -x\exp(-x^2/c_3^2)$, and $f(x) = x\exp(-x^2/c_3^2)$ are considered as the input functions of the system corresponding to **Figures 5-8**, respectively, in which $c_1$=0.001, $c_2$=0.01, and $c_3$=0.01. The input field profiles, the simulated output signals, and the ideal analytical ones are demonstrated in **Figures 5-8 (b), (d)**. We should remark that for ease of presentation, suitable phase factors $\exp(i\varphi)$ have been involved in the output profiles as the multiplicative coefficients to ensure a pure real field at the observation plane. As can be seen, the proposed space-time coding processing system reveals a good performance by exposing the reflected fields whose spatial variations are in perfect agreement with the exact responses.

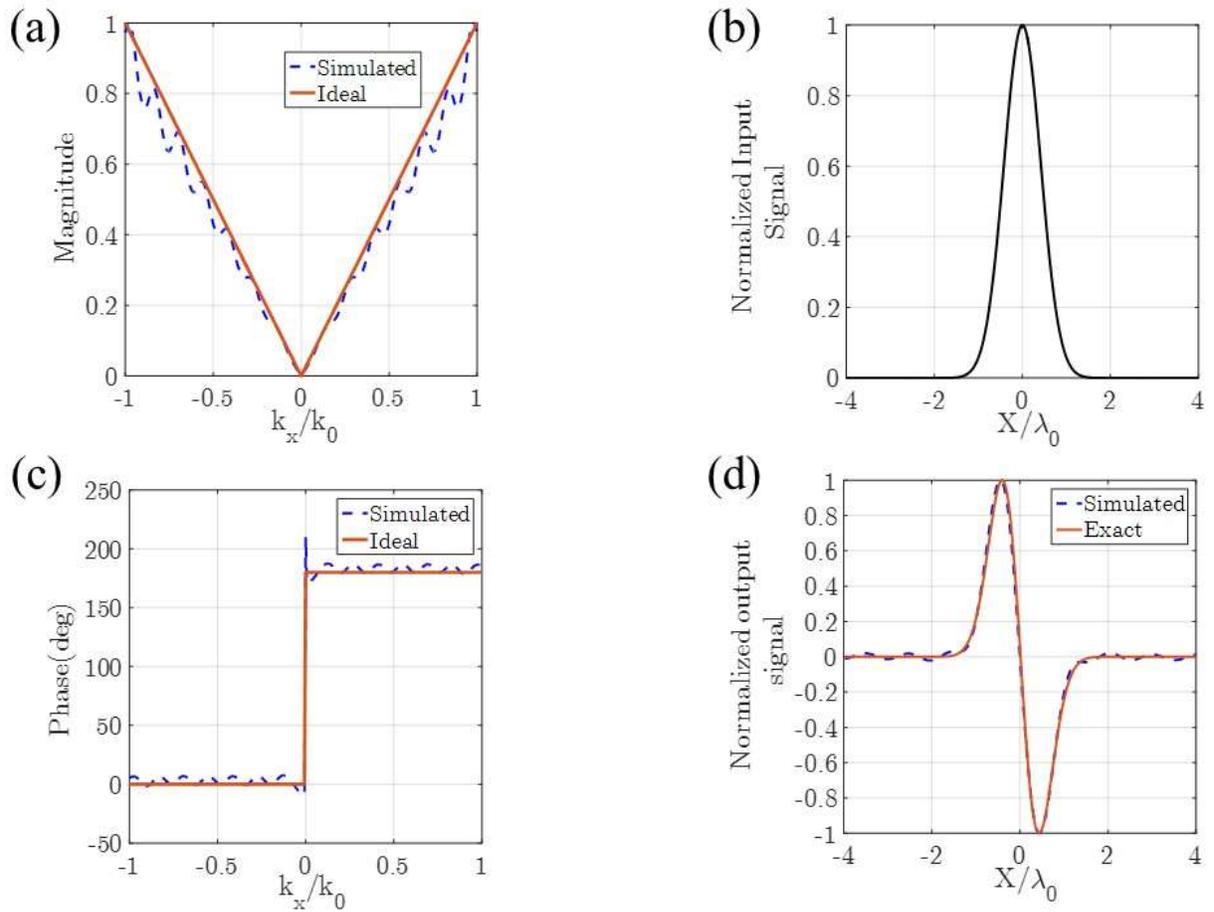

**Figure 5.** A comparison between the simulated and ideal results for (a) amplitude and (c) phase of the transfer function belonging to the 1st-order spatial differentiation and (d) the output signal corresponding to (b) a Gaussian input signal.

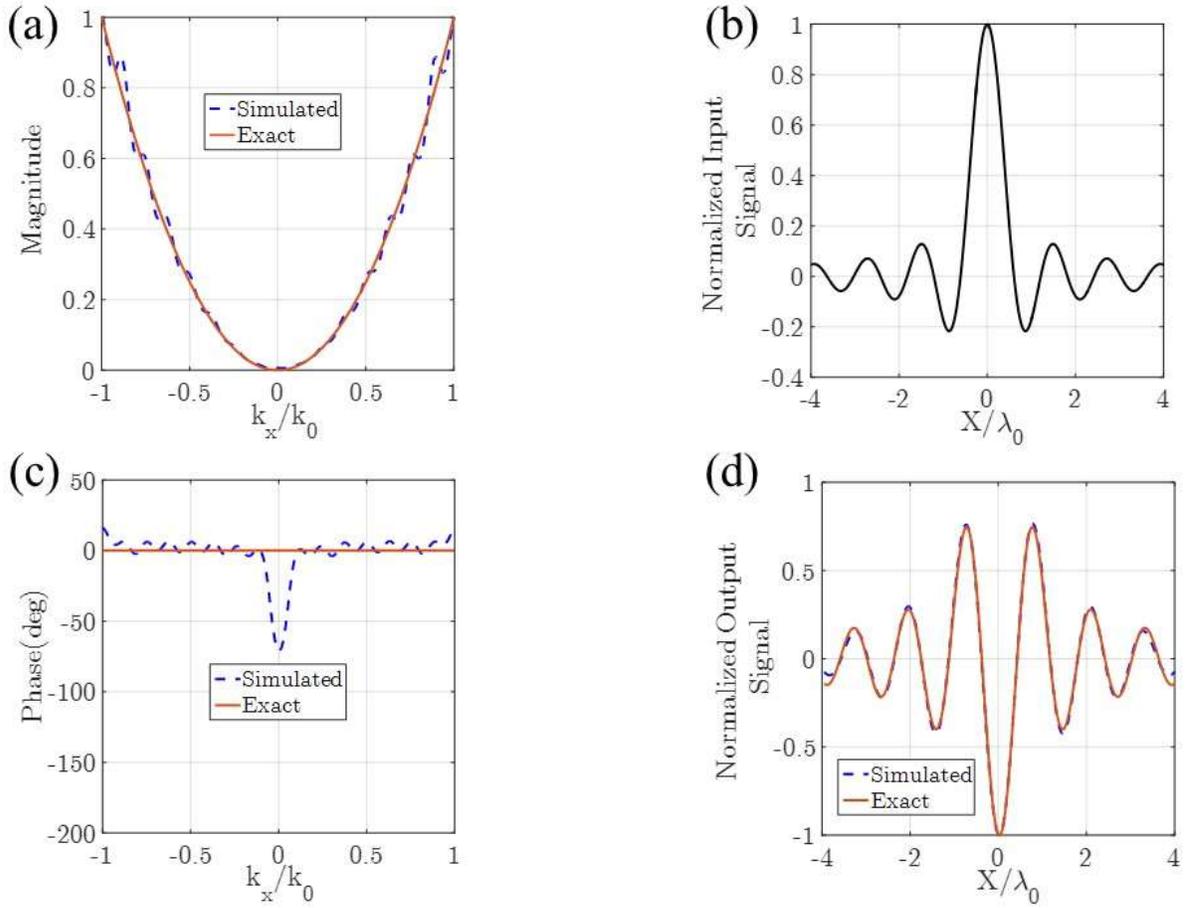

**Figure 6.** A comparison between the simulated and ideal results for (a) amplitude and (c) phase of the transfer function belonging to the 2nd-order spatial differentiation and (d) the output signal corresponding to (b) a sinc-shape input signal.

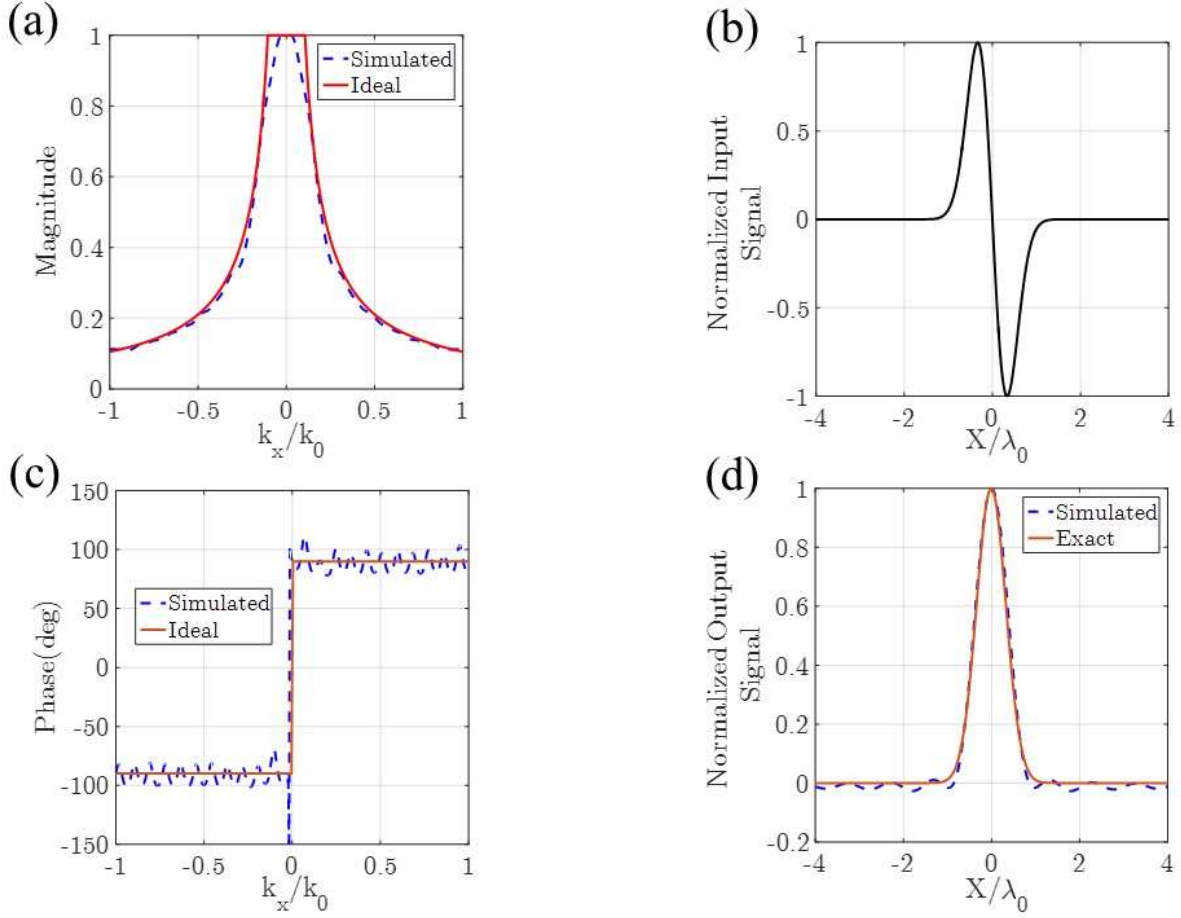

**Figure 7.** A comparison between the simulated and ideal results for (a) amplitude and (c) phase of the transfer function belonging to the $1^{st}$-order spatial integration and (d) the output signal corresponding to the (b) input signal $f(x) = -x\exp(-x^2/c_3^2)$.

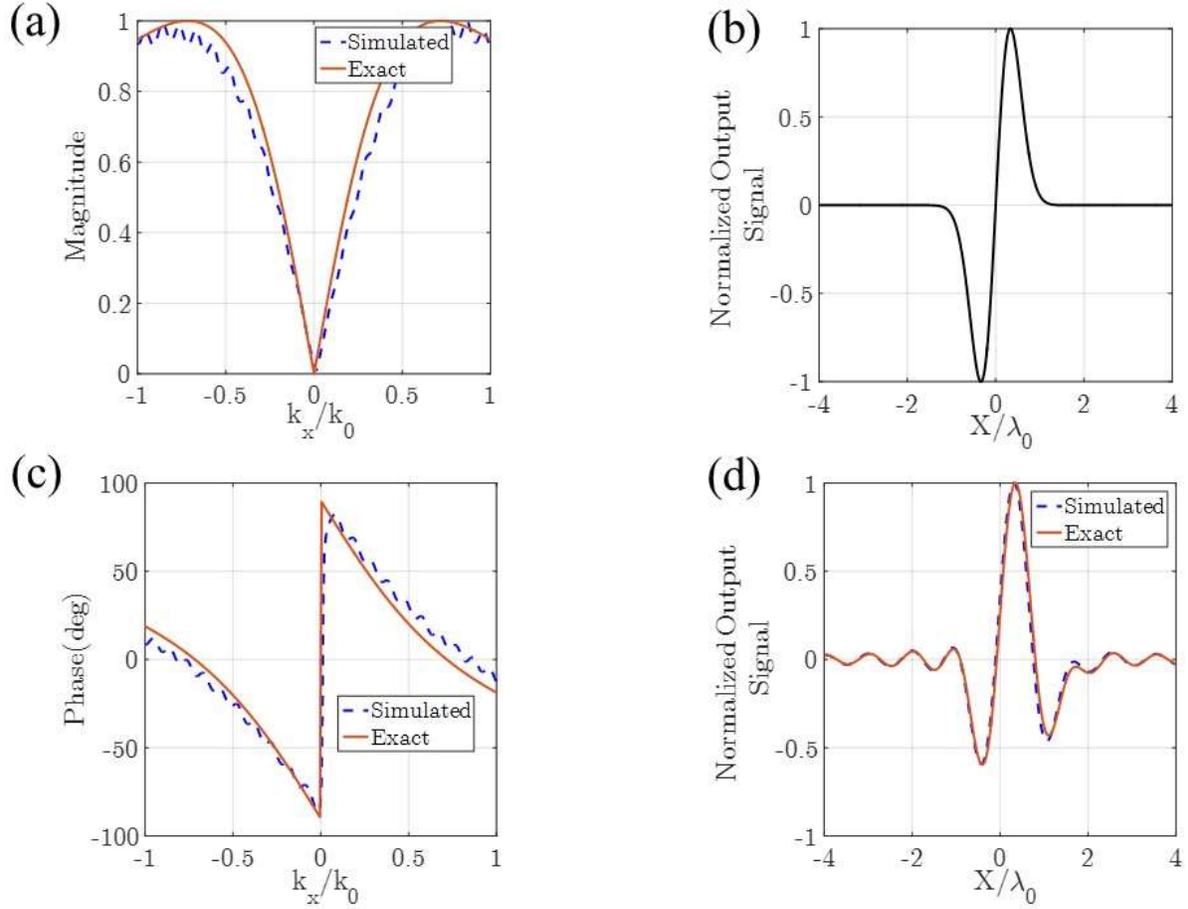

**Figure 8.** A comparison between the simulated and ideal results for (a) amplitude and (c) phase of the transfer function belonging to the integro-differential equation solving operator and (d) the output signal corresponding to the (b) input signal $f(x) = x\exp(-x^2/c_3^2)$.

At the end, we intend to demonstrate the potential applications of the proposed 1D differentiator in image processing, where edge detection is known as a fundamental step [22]. This terminology enables the extraction of boundaries between two regions with different texture characteristics. In the 1D differentiated image being modulated along the x-direction, edges along the horizontal direction will have higher pixel intensity levels than those surrounding them, yielding an instrumental directional selectivity. The simulated transfer functions are exploited to perform 1D spatial filtering on "IRAN CHINA" words along with three rectangular shapes (**Figure 9a**) and "Butterfly" picture (**Figure 9c**) as two different image fields shinning the space-time metasurface. The normalized output fields are achieved, as shown in **Figures 9b, d,** which successfully expose all outlines of the incident images in the horizontal orientations with the same intensity. Although the proof-of-concept simulations are related to 1D space-time metasurface configurations, we can simply extend them into 2D signal processing schemes by programming each meta-atom in an arbitrary row and column, independently, to implement the required spatial Fourier content of the transfer function along either single or both directions.

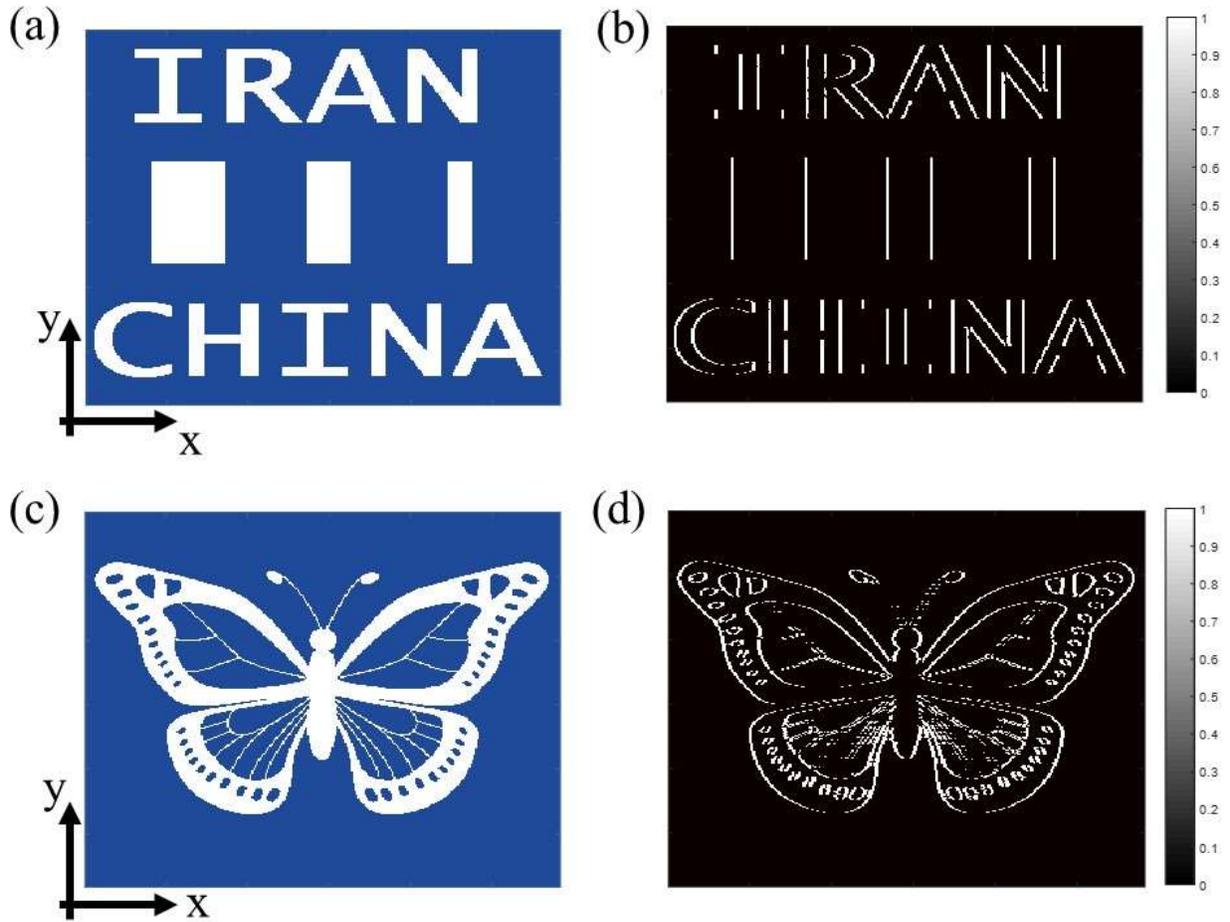

**Figure 9.** Illustration of the edge detection capability of the proposed 1D space-time metasurface processor upon accomplishing the $1^{st}$-order spatial differentiation. (a), (b) The input and (b), (d) output images. As can be seen, all outerior boundaries along the horizontal direction have been successfully extracted.

## 5. Conclusion

In this work, a re-programmable space-time digital metasurface processor was elaborately designed to be utilized in an analog computing platform based on the metasurface approach. Through applying proper time-domain coding signals, the space-time digital metasurface is empowered to dynamically realize the required phases and amplitudes of the transfer function associated with the operator of choice. Several illustrative examples were presented to demonstrate the versatility of the proposed metasurface processor in performing diverse mathematical operations and functionalities such as spatial differentiation, spatial integration, solving integro-differential equations, and edge detection in a real-time manner. The proposed space-time coding strategy takes a great step forward in developing re-programmable wave-based signal processors with offering great versatility in the operations.